\documentclass[american, aps,pra, twocolumn,  reprint ]{revtex4-1}
\usepackage[unicode=true,pdfusetitle, bookmarks=true,bookmarksnumbered=false,bookmarksopen=false, breaklinks=false,pdfborder={0 0 0},backref=false,colorlinks=false] {hyperref}
\hypersetup{ colorlinks,linkcolor=myurlcolor,citecolor=myurlcolor,urlcolor=myurlcolor}
\usepackage{braket,colortbl,cleveref,amsthm,amsmath,amssymb,txfonts}
\definecolor{myurlcolor}{rgb}{0,0,0.9}
\usepackage{graphics,graphicx}
\usepackage{color}
\usepackage{graphicx}
\usepackage[format = plain,labelfont = bf,up, textfont = normal , up, justification =raggedright, singlelinecheck =false]{caption}
\usepackage{subcaption}
\usepackage{rotating}
\usepackage{tabularx}
\usepackage{amsmath}
\DeclareMathOperator{\tr}{Tr}
\usepackage{graphicx}
\usepackage[utf8x]{inputenc}
\usepackage{color}
\usepackage{amsmath}
\usepackage{braket}
\usepackage{latexsym}
\usepackage{amssymb}
\usepackage{amsthm}
\usepackage{bm}
\usepackage{graphics,epstopdf}
\usepackage{color}\usepackage{amsmath}
\usepackage{enumitem}
\usepackage{fmtcount}
\usepackage{booktabs}
\usepackage{epsfig}
\theoremstyle{plain}
\def\bea{\begin{eqnarray}}
\def\eea{\end{eqnarray}}
\def\ba{\begin{array}}
\def\ea{\end{array}}

\def\ket{\rangle}
\def\bra{\langle}
\def\beq{\begin{equation}}
\def\eeq{\end{equation}}

\begin{document}
 
\title{Generating steady quantum coherence and magic through an autonomous thermodynamic machine by utilizing a spin bath}

\author{Chiranjib Mukhopadhyay}
\address{Quantum Information and Computation Group, Harish Chandra Research Institute, Homi Bhabha National Institute, Allahabad 211019, India}
\email{chiranjibmukhopadhyay@hri.res.in}

\begin{abstract} 
\noindent  Generation of steady quantumness in the presence of an environment is of utmost importance if we are to build practical quantum devices. We propose a scheme of generating steady coherence and magic in a qubit system attached to a heat bath at a certain temperature through its interaction with another qubit system attached to a spin bath. Coherence generation in the reduced qubit is always possible in this model. The steady coherence in the reduced qubit attached to the heat bath may be used to enhance the subsequent transient performance of a quantum absorption refrigerator. For the case of generation of magic, which is the quantum resource responsible for implementation of gates which are not simulable via stabilizer computation,  we show that there exists a critical temperature of the heat bath beyond which it is not possible to create magic in the reduced qubit attached to the heat bath. Below the critical temperature, the strength of interaction between the qubits must lie within a certain region for creation of magic. We further note that by increasing the strength of coupling of the second qubit to the spin bath, typified by the reset probability, keeping the coupling strength of the first qubit to the heat bath, it is possible to increase the critical temperature of the heat bath for creation of magic. 

 
\end{abstract}

\maketitle

\section*{Introduction}
Quantum theory allows for the accomplishment of information processing and computational tasks \citep{nielsen,wilde} which are impossible or difficult to achieve through the classical paradigm. However, the quantum properties which underlie these operational advantages are notoriously fragile. In particular, inevitable contact with an environment may lead to decoherence and loss of quantum features. This stands in the way of realizing quantum devices in practical situations.  Since completely sealing off quantum systems from the environment is very difficult, we may alternately ask - can we somehow use the environment as an ally instead of an impediment \citep{friend, breuer_book} ? This broad area of research has received renewed attention in recent years with the advent of bath-engineering techniques as well as works on non-Markovian environments \citep{be1, be2, petruccione, nm1,nm2, our1, our2}. One particular realization in the recent past is of the fact that apart from heat baths, baths such as spin baths can be used to overcome the Landauer erasure energy cost \citep{landauer,landaueram, lostag}, although a corresponding cost has to be paid in terms of angular momentum. Non-thermal baths may also offer better efficiency for various thermodynamic designs \citep{nonthermal1,nonthermal2,nonthermal3,nonthermal4,nonthermal5}
.

Various autonomous quantum thermal machines have been recently envisaged for thermodynamic tasks such as  refrigeration \citep{popescu,jpa, brunner,levy,brask,mitchison, alonso, correa, iontrap, abs3, abs4, abs5, bsocr} as well as rectification \citep{diode, transistor}, thermometry \citep{hofer} or in setting up an autonomous clock \citep{clock}. With some modification, such a model was recently proposed \citep{brunner_ent_gen} where entanglement may be obtained in the steady state configuration of the joint state of the qubits which comprise the thermal machine. However, the reduced state of individual qubits in such models \citep{popescu} are still diagonalized in the energy eigenbasis and consequently, have no quantum signature in the form of quantum coherence \citep{baumgratz, colloquium} or \emph{magic} \citep{njp_magic,goursanders}. The latter is especially important from an operational perspective as it is defined to be the property by virtue of which an ancilla state can help in universal quantum computation via \emph{stabilizer codes}, a restricted and classically simulable subset of quantum operations on a larger Hilbert space \citep{nielsen}.

In this work, we show, in a self-contained model, how to impart quantum properties to the steady state of a qubit system interacting with a thermal bath utilizing an angular momentum bath interacting with another qubit. We propose the setup and using a simple reset model, explicitly find the steady state configuration. This enables us to observe how non-classicality in the form of quantum coherence and magic builds up in the steady state. Simply equilibrating a qubit in the angular momentum bath instead of the heat bath may yield coherence in the energy eigenbasis, but may not yield magic. However, in the proposed setup, we shall show that the reduced qubit in its steady state  may indeed have non-zero magic, i.e., be useful as ancilla for non-classical gate implementation.  

The structure of the paper is as follows. In section \ref{sec1}, we introduce the setup analyzed in the present work. This is followed by section \ref{sec2} where we discuss the  generation of steady coherence in the steady state of the reduced qubit attached to the heat bath.  Section \ref{sec3} is devoted to the analysis of generation of magic in the steady state of the qubit for various parameter regimes. This is followed by a conclusion and discussion about possible future aveneues of work.

\section{Model}
\label{sec1}
\begin{figure}
\includegraphics[width = 0.5\textwidth, keepaspectratio]{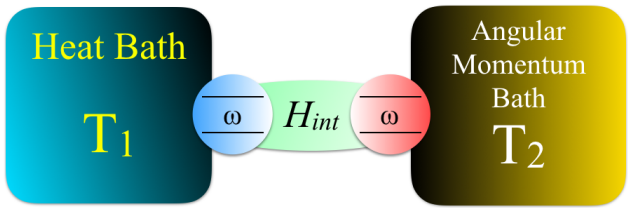}
\caption{Schematic diagram of model used in the work.}
\label{schematic}
\end{figure}

The concept of canonical ensemble in statistical mechanics usually refers to the situation where a physical system exchanges energy with the environment to equilibriate. According to Jaynes' MaxEnt principle, it can be shown that the population of equilibrium density matrix of the state which maximizes the information theoretic entropy for a given amount of average energy follows the Gibbs distribution with a potential-like parameter $T$, which we term as \emph{temperature}. However, instead of energy, we can envision a situation where the system exchanges spin angular momentum along a specified direction, say $\vec{k}$. In this situation, again maximizing the information theoretic entropy for a given amount of average spin angular momentum along $\vec{k}$-direction yields an equilibrium state which follows the Gibbs distribution with a potential-like parameter which plays a role similar to the role of temperature for heat baths. In this paper, we shall loosely call this parameter as temperature of the angular momentum bath. However we warn the reader, that this \emph{temperature} in this context, is to be understood as something different from the way temperature is used in the $usual$ sense for thermal baths. It is natural to wonder about the theoretical as well as experimental basis for assuming such baths. Theoretically, the motivation comes chiefly from Vaccaro and Barnett's \citep{landaueram} pioneering work, showing such baths can give rise to Landauer erasure without energy cost. More recent recent resource theoretic works \citep{lostag, guryanova, oppenheim} consider even more general kinds of baths with any number of conserved charges, of which the bath proposed above is a very special case. A recent work on cyclic thermal machines between a thermal and a spin reservoir has also appeared in this connection \citep{wright}. However, practical realization of these baths seem to be rather difficult, as has been pointed out, for example, in \citep{physics}. Hence, we reserve comment on the actual practical realization of our model.  In this context, we also note for clarity that the spin bath is $not$ a thermal resource in the usual sense, and hence the quantumness generation procedure outlined in this work, although autonomous, is not altogether thermal. The reader may compare and contrast this approach with another recent work \citep{manzano}. 
\\

Let us now introduce the setup in Fig. \ref{schematic}.  The first qubit is immersed in a heat bath of temperature $T_1$, where the energy eigenbasis is along the $z$ direction. The second qubit is immersed in a spin bath of temperature $T_2$, where the spin angular momenta along $x$ direction are exchanged. The Hamiltonian corresponding to the first qubit is \beq H_{1} =  \frac{1}{2}\omega_1 |1\ket\bra 1| \ , \eeq and the  Hamiltonian corresponding to the second qubit is \beq H_2 = \frac{1}{2} \omega_2 |1\ket\bra 1|  \ .\eeq  We also assume an energy swapping interaction $H_{int} = g \left( |01\ket\bra 10| + |10\ket\bra 01| \right)$ between the two qubits. We assume the resonance condition $\omega_1 = \omega_2 = \omega$. In subsequent calculations, we shall assume $\omega =1$ without loss of generality. Thus the collective Hamiltonian reads as \beq H =  H_{1} \otimes \mathbb{I} + \mathbb{I} \otimes H_2 + H_{int}\eeq

The thermal state of a particle, immersed in the heat bath of inverse temperature $\beta_1 = 1/ T_1$, now reads as   \beq \tau_1 = \frac{1}{1 + e^{-\beta_1}} |0\ket\bra 0| + \frac{e^{-\beta_1}}{1 + e^{-\beta_1}} |1\ket\bra 1|.   \eeq The corresponding equilibrium state of a particle, immersed in the spin bath of inverse temperature $\beta_2 = 1/ T_2$,  is given by \beq  \tau_2 = \frac{1}{1 + e^{-\beta_2}} |+\ket\bra +| + \frac{e^{-\beta_2}}{1 + e^{-\beta_2}} |-\ket\bra -|  \eeq

During each small time interval $\delta t$ of the dynamics,  one of the qubits of the two qubit state $\rho_{12} (t)$ can thermalize back to its respective equilibrium configuration (that is, $\tau_1$ for the first qubit and $\tau_2$ for the second qubit) with probabilities $p_1$ and $p_2$ respectively. We assume that the probability of both the qubits equilibrating in $\delta t$ interval is negligible. Thus, the master equation for the two qubits read as the following 

\beq \frac{d\rho_{12}(t)}{dt} = -i [H,\rho_{12}] + \sum_{i} p_{i} \left( \tau_{i}\otimes \tr_{i} \rho_{12} (t) - \rho_{12} (t) \right) \label{qme} \eeq

\noindent In general, the steady state of a qubit immersed in a bath and oblivious to any other system, and the the state that the same qubit wants to revert to, \emph{while} interacting with another system, may  be different. This may be especially prominent if the qubit in question is  coupled very strongly with the other qubit, when compared with the coupling with the bath. Thus, we shall restrict ourselves to the weak interaction strength, i.e., $g$ being small, when usng the above master equation.  The steady state $\rho_{12}^{steady}$ is obtained by solving for vanishing right hand side of the evolution master equation \eqref{qme}. Since a general two-qubit density matrix has fifteen real parameters, this implies solving a system of linear equations with fifteen variables. We quote the general solution in the supplementary material. However, the general expression for the  steady state is algebraically quite cumbersome - therefore we shall  state  and use simplifying assumptions in the rest of the paper.  

\section{Analysis of quantum coherence generation in the reduced qubit}
\label{sec2}
\begin{figure}
\includegraphics[width = 0.23\textwidth, keepaspectratio]{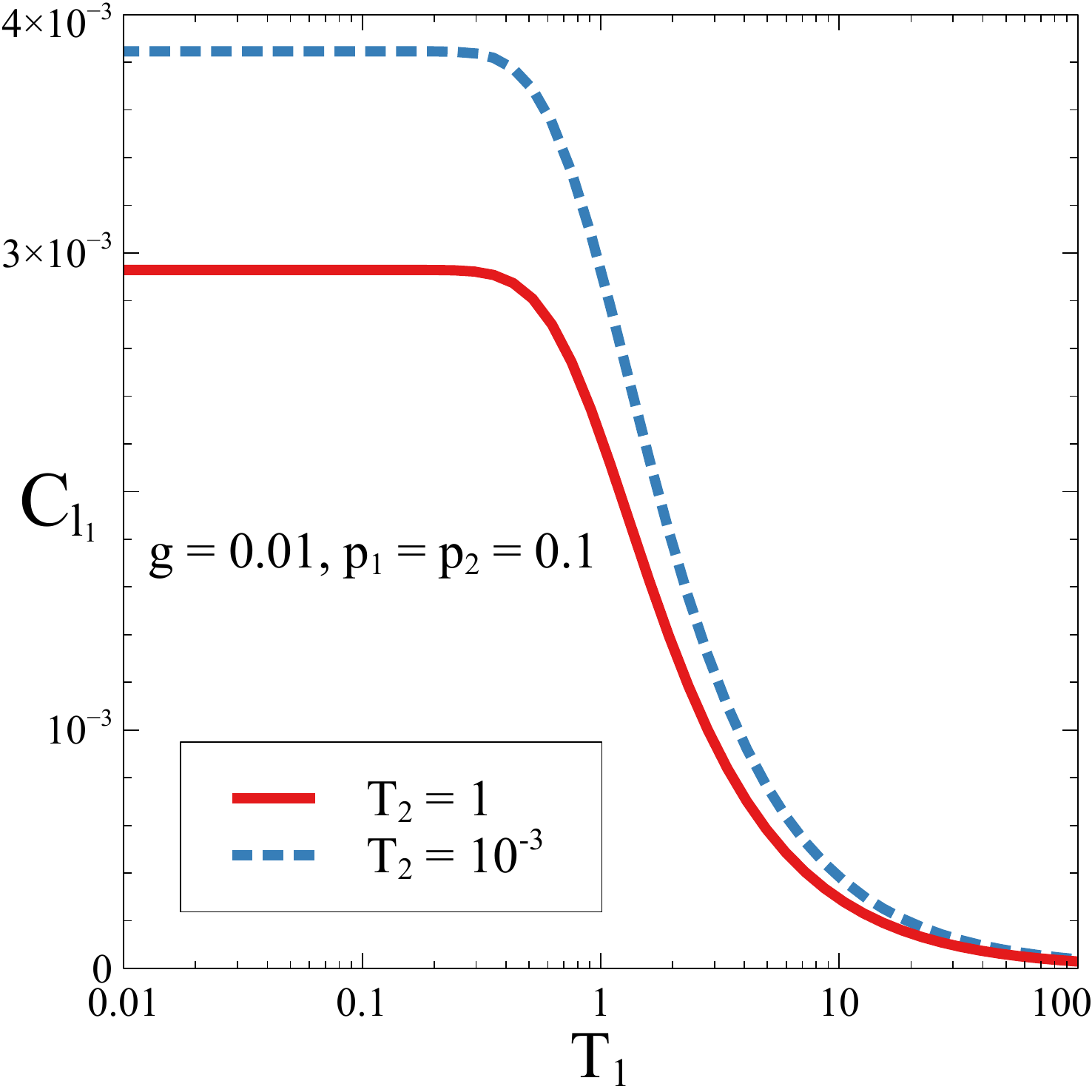} \quad
 \includegraphics[width = 0.23\textwidth, keepaspectratio]{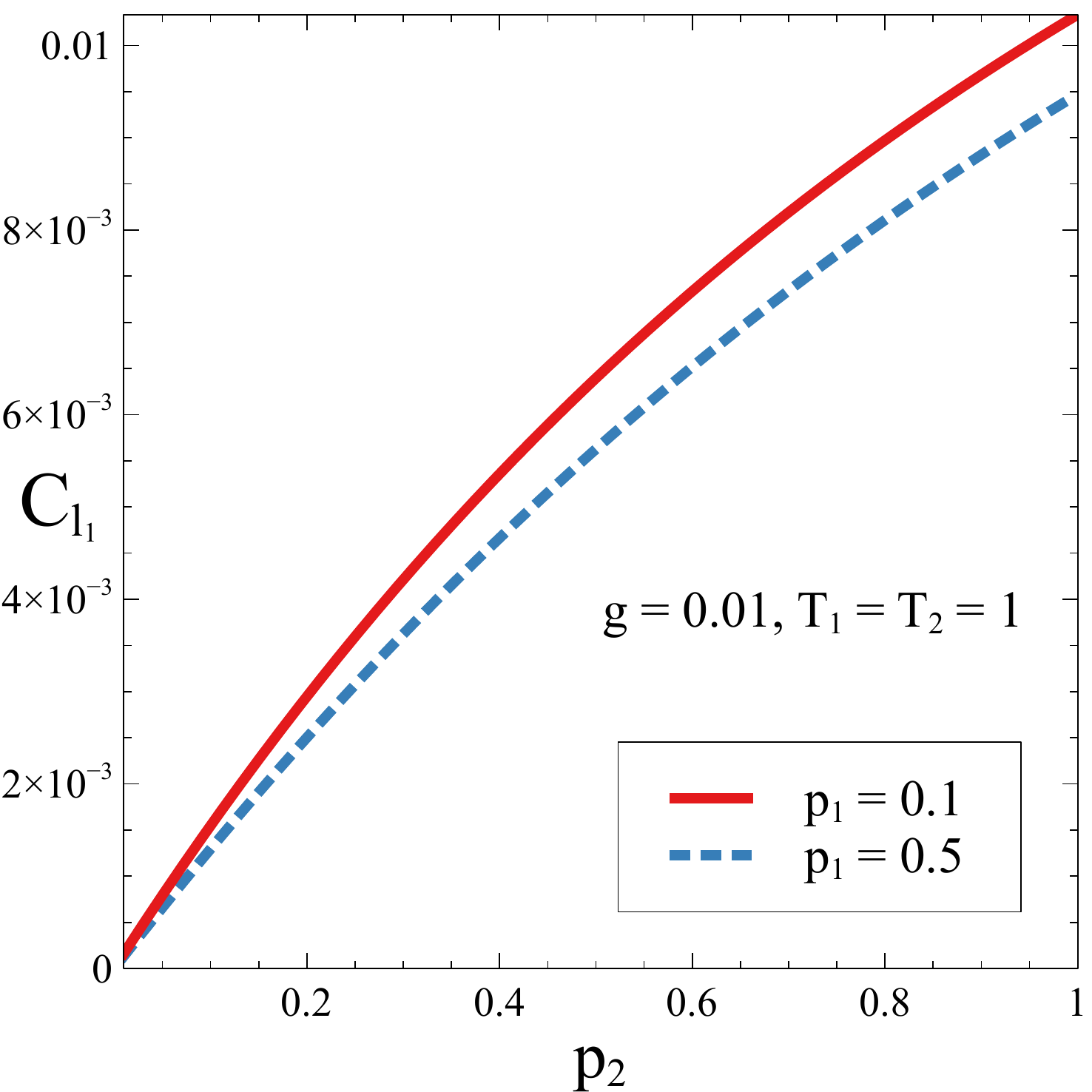} 
 \caption{Dependence of the steady state coherence of the first qubit on the heat bath temperature $T_1$ (left) and the reset probability for the spin bath $p_2$ (right).}
\end{figure}

Quantumness in the form of superposition with respect to a fixed basis has been recently formally quantified through the resource theory of quantum coherence and linked with various other operational quantum resources such as entanglement \citep{uttament}, general quantum correlations \citep{mile,yao,killoran}, quantum fisher information \citep{kwek} or stabilizer computation \citep{chiru9}. From the perspective of autonomous quantum thermal machines, initial coherence is a resource for augmenting the performance of an absorption refrigerator \citep{brask,mitchison, avijitsreetama}. The inverse problem of creating coherence in finite dimensional systems using thermal resources has also attracted recent attention \citep{manzano}. As we shall show below, the reduced steady state of the qubit attached to the heat bath is coherent in the energy eigenbasis. Thus, when the steady state is reached in our setup, if we simply strip the other components of the present model (except the heat bath and the attached qubit) away and replace them with the hot and cold heat baths of the quantum absorption refrigerator setup, we can benefit from the initial coherence in the absorption refrigerator setup.

From the general steady state solution furnished in the supplementary material, if one performs a perturbative expansion for small interaction strength $g$, the $l_1$-norm of coherence in the reduced qubit attached to the heat bath reads as \beq C_{l_{1}}  = \frac{4g p_2}{\sqrt{(1 + 4 p_1^2)(1 + 4 p_2^2)}} \left| \tanh \left( \frac{1}{2 T_1} \right) \tanh \left( \frac{1}{2 T_2} \right) \right|  + \mathcal{O} (g^2) \eeq

The first observation is that increased thermalization probability $p_1$ leads to a decrease of the steady coherence. The second observation is that for small thermalization probability $p_2$, increasing it also increases the magnitude of steady coherence. However, as we go on increasing the reset probability $p_2$, the magnitude of steady coherence asymptotically reaches a maximum. Regarding the bath temperatures, we observe that the magnitude of steady coherence is increased if the bath temperatures are low. 

\section{Generation of magic in the reduced qubit}
\label{sec3}

Many fault-tolerant quantum algorithms use the so called $stabilizer$ $operations$, i.e., unitary gates  and measurements chosen from a specific set. It can be shown via the Gottesmann-Knill theorem, that these set of operations are efficiently simulable via classical means. Thus, for universal quantum computation, if one only allows for $stabilizer$ operations, one must introduce additional ancilla states along with the original system. Stabilizer operations may then be peformed over the larger Hilbert space consisting of the original system plus the  ancilla to effectively implement non-stabilizer operations on the actual system. In order to facilitate non-stabilizer operations on the original system, the ancilla states must lie outside the convex hull of  pure states, which are known as $stabilizer$ $states$. States which satisfy this property are defined to be endowed with \emph{magic}. Thus, just as quantum entanglement is the operational resource underlying the superiority of quantum communication protocols, \emph{magic} is the resource for classically non-simulable gate implementation \citep{njp_magic, contextuality_supplies_magic}. Thus, creation of magic in a quantum system is vital for quantum technology. 
\begin{figure}
\includegraphics[width = 0.4\textwidth, keepaspectratio]{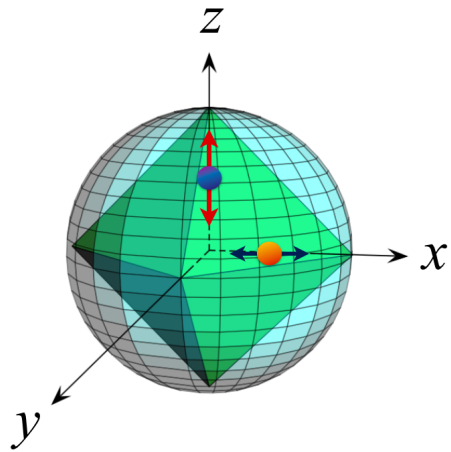}
\caption{If a qubit state equilibrates on the $z$-axis of the Bloch sphere (blue blob) or on the $x$-axis of the Bloch sphere (golden blob) - the state lies within the stabilizer polytope. The first scenario is associated with the thermal state of the heat bath, the second with the equilibrium state  of the angular momentum bath depicted in our model. }
\label{magic_bloch}
\end{figure}
Indeed, if we simply immerse a qubit to the heat bath, it thermalizes at an equilibrium state which lies on the $z$-axis of the Bloch sphere, i.e., always within the stabilizer polytope. More interestingly, if a qubit is immersed in the spin bath described above, then the steady state lies on the $x$-axis of the Bloch sphere, i.e., again within the stabilizer  polytope, although it may be coherent in the energy eigenbasis. See Fig. \ref{magic_bloch} for an illustration. Nonetheless, we shall now show that the magic can be indeed imparted in the steady state of the qubit attached to the heat bath through our setup. 

In the qubit case, the states which can \emph{not} be used as ancilla to implement classically non-simulable gates, lie inside the convex polytope formed by the eigenvectors of the mutually unbiased operators $\sigma_x, \sigma_y, $ and $\sigma_z$. Any state outside this so called $stabilizer$ polytope is said to possess magic. In terms of the Bloch vector $\vec{r} = (r_x,r_y,r_z)$ of a quantum state, the condition for the state lying within the stabilizer polytope is when all the following inequalities are simultaneously met \citep{goursanders}. \beq -1 \leq r_x \pm r_y \pm r_z \leq 1 .\label{mag_cond}\eeq For qutrit and other higher prime power dimensional states,  the negativity of the discrete Wigner function is an analytically computable  magic monotone. However, for qubits, the situation is less fortunate. While magic monotones like relative entropy of magic \citep{njp_magic}, robustness of magic \citep{contextuality_supplies_magic} or SDP based measures  \citep{goursanders} indeed exist in the qubit case - they are not amenable to simple analytical calculations. Thus, we would only investigate the condition for the existence of magic. 
 \begin{figure}
\includegraphics[width = 0.23\textwidth, keepaspectratio]{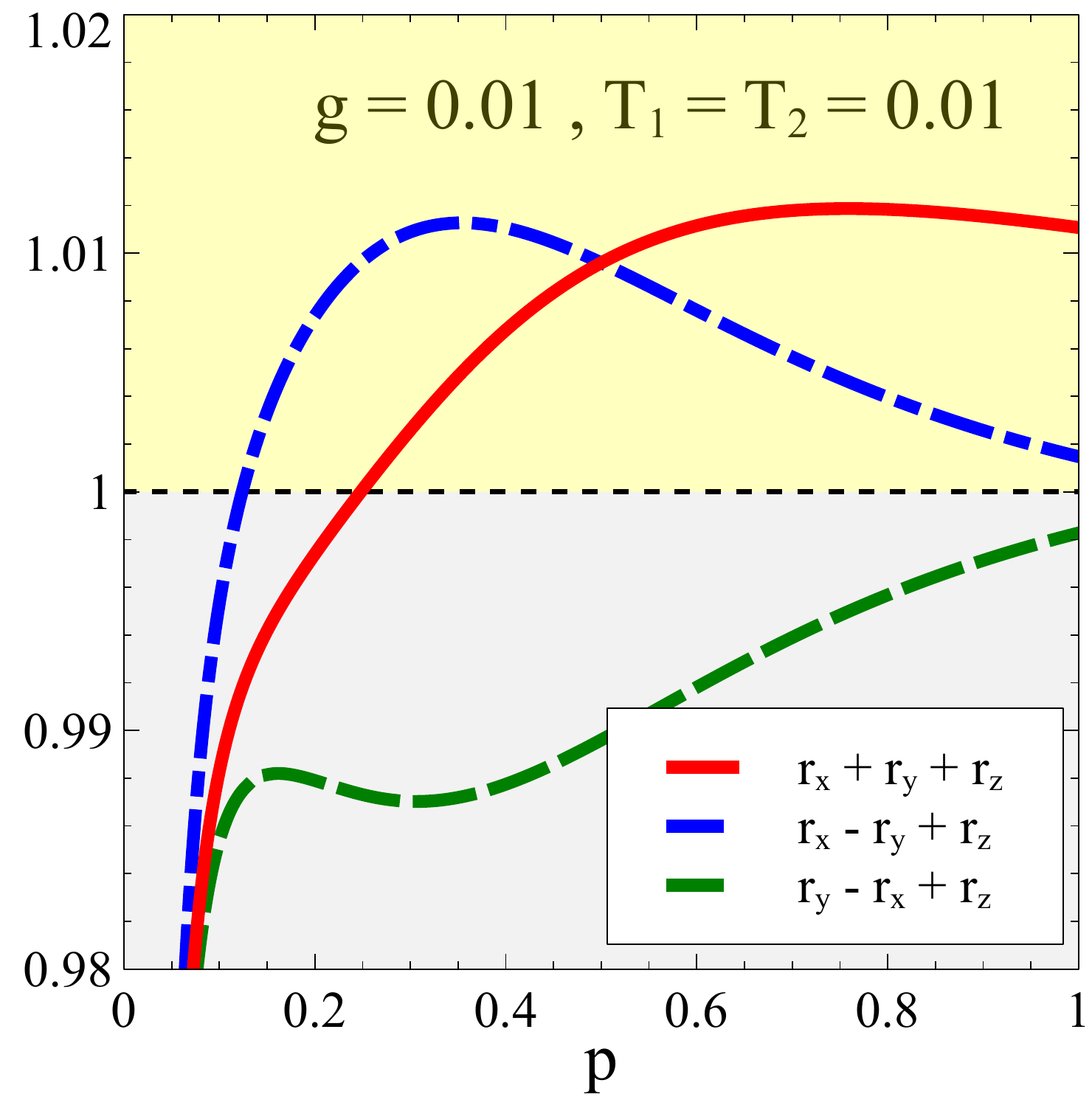} \quad
\includegraphics[width = 0.23\textwidth, keepaspectratio]{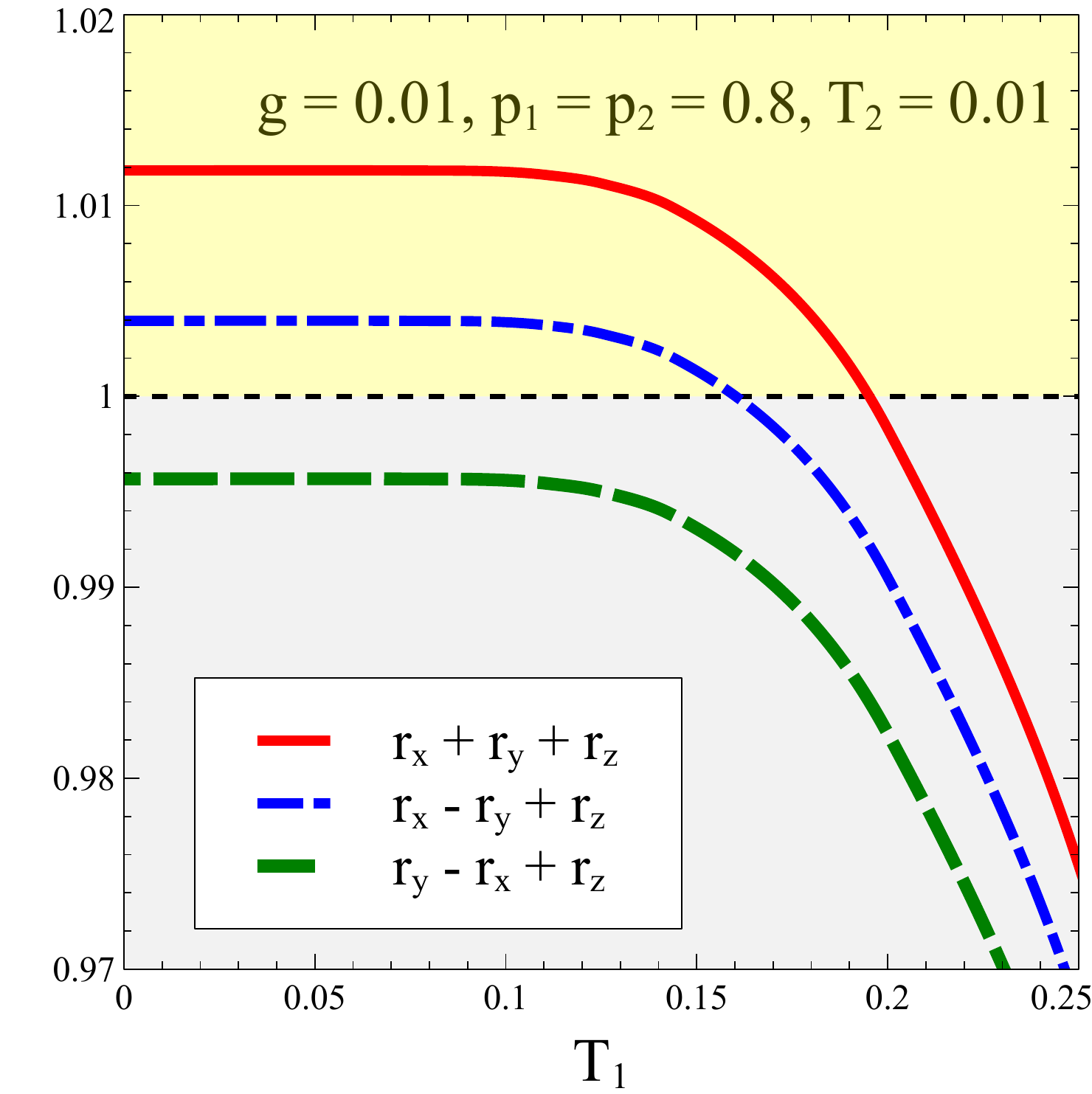} \quad
\caption{Values of linear functions of Bloch vector (cf. \eqref{mag_cond} of the reduced qubit attached to the heat bath vs. the reset probability $p_1 = p_2 = p$ (left) and the heat bath temperature $T_1$ (right). Any one of the curves falling in the pale yellow region indicates the presence of magic. If all three lines lie in the light grey region, that indicates the state is within the stabilizer polytope. }
\label{exact_plots}
\end{figure}
Fig \ref{exact_plots} depicts the results for the exact steady-state solution, which indicate that the above quantities in \eqref{mag_cond} can indeed exceed unity and thus create magic in the reduced qubit attached to the heat bath. Fig \ref{exact_plots} also allows us to observe that as we go on increasing the heat bath temperature, the value of the quantities in \eqref{mag_cond} eventually stop exceeding unity. Thus there seems to be a critical temperature associated with the heat bath above which magic creation may not be possible in the reduced qubit attached to the heat bath. 
\\
Proving the above results from the full steady state solution, which is algebraically messy, is quite challenging. Instead, as a way of simplification, we will follow a perturbative approach, inspired by the fact that the quantum master equation \eqref{qme} holds true if the interaction strength $g$ is weak. From the general expression for the Bloch vectors of the qubit attached to the heat bath of temperature $T_1$, we may write down the leading order terms for the perturbation expansion for small $g$ as -
\begin{widetext}
\bea 
r_x + r_y + r_z = \tanh \left(\frac{1}{2 T_1} \right)  \left[ 1 + 4g \frac{p_2 \left( -1 + 2 p_2 + 2 p_1 + 4 p_1 p_2 \right) \tanh \left( \frac{1}{2 T_2} \right)}{(1 + 4 p_1^2)(1 + 4 p_2^2)}  + 2 g^2 \frac{\left[-1 - 2 p_2^2 - 4 p_1 \left(p_1 + 4 p_1 p_2^2 + 2 p_2^3 \right)  + \frac{4 p_2^2 (4 p_1 p_2 -1)}{1 + \cosh(\frac{1}{T_2})}\right] }{p_1 (p_1 + p_2) (1 + 4 p_1^2)(1 + 4 p_2^2)} \right]  \nonumber \\
r_x  - r_y + r_z = \tanh \left(\frac{1}{2 T_1} \right)  \left[ 1 + 4g \frac{p_2 \left( 1 + 2 p_2 + 2 p_1 - 4 p_1 p_2 \right) \tanh \left( \frac{1}{2 T_2} \right)}{(1 + 4 p_1^2)(1 + 4 p_2^2)}  + 2 g^2 \frac{\left[-1 - 2 p_2^2 - 4 p_1 \left(p_1 + 4 p_1 p_2^2 + 2 p_2^3 \right)  + \frac{4 p_2^2 (4 p_1 p_2 -1)}{1 + \cosh(\frac{1}{T_2})}\right] }{p_1 (p_1 + p_2) (1 + 4 p_1^2)(1 + 4 p_2^2)} \right]  \nonumber \\
-r_x + r_y + r_z =  \tanh \left(\frac{1}{2 T_1} \right)  \left[ 1 + 4g \frac{p_2 \left( -1 - 2 p_2 - 2 p_1 + 4 p_1 p_2 \right) \tanh \left( \frac{1}{2 T_2} \right)}{(1 + 4 p_1^2)(1 + 4 p_2^2)}  + 2 g^2 \frac{\left[-1 - 2 p_2^2 - 4 p_1 \left(p_1 + 4 p_1 p_2^2 + 2 p_2^3 \right)  + \frac{4 p_2^2 (4 p_1 p_2 -1)}{1 + \cosh(\frac{1}{T_2})}\right] }{p_1 (p_1 + p_2) (1 + 4 p_1^2)(1 + 4 p_2^2)} \right]  \nonumber \\
\eea
\end{widetext}

Let us now concentrate on specific parameter domains to explicitly find out the condition for existence of magic. We begin with the assumption that the temperature $T_2$ is very low and assume further that the reset probabilities $p_1$ and $p_2$ are equal in magnitude and have the value, say,  $p$.  Under these conditions 
\begin{widetext}
\begin{align}
& r_x + r_y + r_z \approx \tanh \left(\frac{1}{2 T_1} \right) \left[1 + 4 g p \frac{4p^2 + 4p -1}{ (1 + 4 p^2)^2 } -  g^2 \frac{1 + 6 p^2  + 24 p^4}{p^2 (1 + 4 p^2)^2}  \right]   
\end{align}
\end{widetext}
Noting that the condition $r_x + r_y + r_z > 1$ is sufficient for the existence of magic in the reduced qubit, we express this condition under the above assumptions as

\begin{align}
 1 + 4 g \frac{p(4p^2 + 4p -1)}{ (1 + 4 p^2)^2 } - g^2 \frac{1 + 6 p^2  + 24 p^4}{p^2 (1 + 4 p^2)^2}  >  \coth \left(\frac{1}{2 T_1} \right)
\end{align}

Let us now designate $f_1=\frac{p(4p^2 + 4p -1)}{ (1 + 4 p^2)^2 }, f_{2}=\frac{1 + 6 p^2  + 24 p^4}{p^2 (1 + 4 p^2)^2} $, and $\lambda = \coth \left(\frac{1}{2 T_1} \right) -1 $, thus the expression above is written as  \beq g^2  -4 g \frac{f_1}{f_2} + \frac{\lambda}{f_2} < 0,  \eeq  which yields the condition \beq \left( g - 2 \frac{f_1}{f_2} \right)^2 < \ 4 \frac{f_1^2}{f_2^2} - \frac{\lambda}{f_2} \label{condition_magic} \eeq Note that, it becomes possible to satisfy the above criteria, only if the right hand side of the above expression is positive. If the reset probabilities are fixed, this implies the existence of a threshold temperature of the hot bath, say $T_{\text{crit}}^{1}$ above which $r_x + r_y + r_z$ can never exceed unity. Similarly analyzing the conditions for $r_x - r_y + r_z$ and $r_y - r_x + r_z$ to exceed unity, give rise to threshold temperatures $T_{\text{crit}}^{2}$, and $T_{\text{crit}}^{3}$ respectively. The actual threshold temperature of the heat bath beyond which magic can not be generated is thus the maximum of these three threshold temperatures, i.e., \beq T_{\text{crit}} = \max \left[T_{\text{crit}}^{1}, T_{\text{crit}}^{2},T_{\text{crit}}^{3}  \right] \eeq where, assuming $g_1 = \frac{p(1 + 4p - 4p^2)}{ (1 + 4 p^2)^2 } $, and $h_1 = \frac{p(4p^2 - 4p -1)}{ (1 + 4 p^2)^2 } $, the critical temperatures are explicitly expressed as \beq T_{\text{crit}}^{1} = \frac{1}{\ln \left(1 + \frac{f_2}{2 f_1^2}\right)}, T_{\text{crit}}^{2} = \frac{1}{\ln \left(1 + \frac{f_2}{2 g_1^2} \right)}, T_{\text{crit}}^{3} = \frac{1}{\ln \left(1 + \frac{f_2}{2 h_1^2} \right)} \eeq

\begin{figure}
\includegraphics[width = 0.23\textwidth, keepaspectratio]{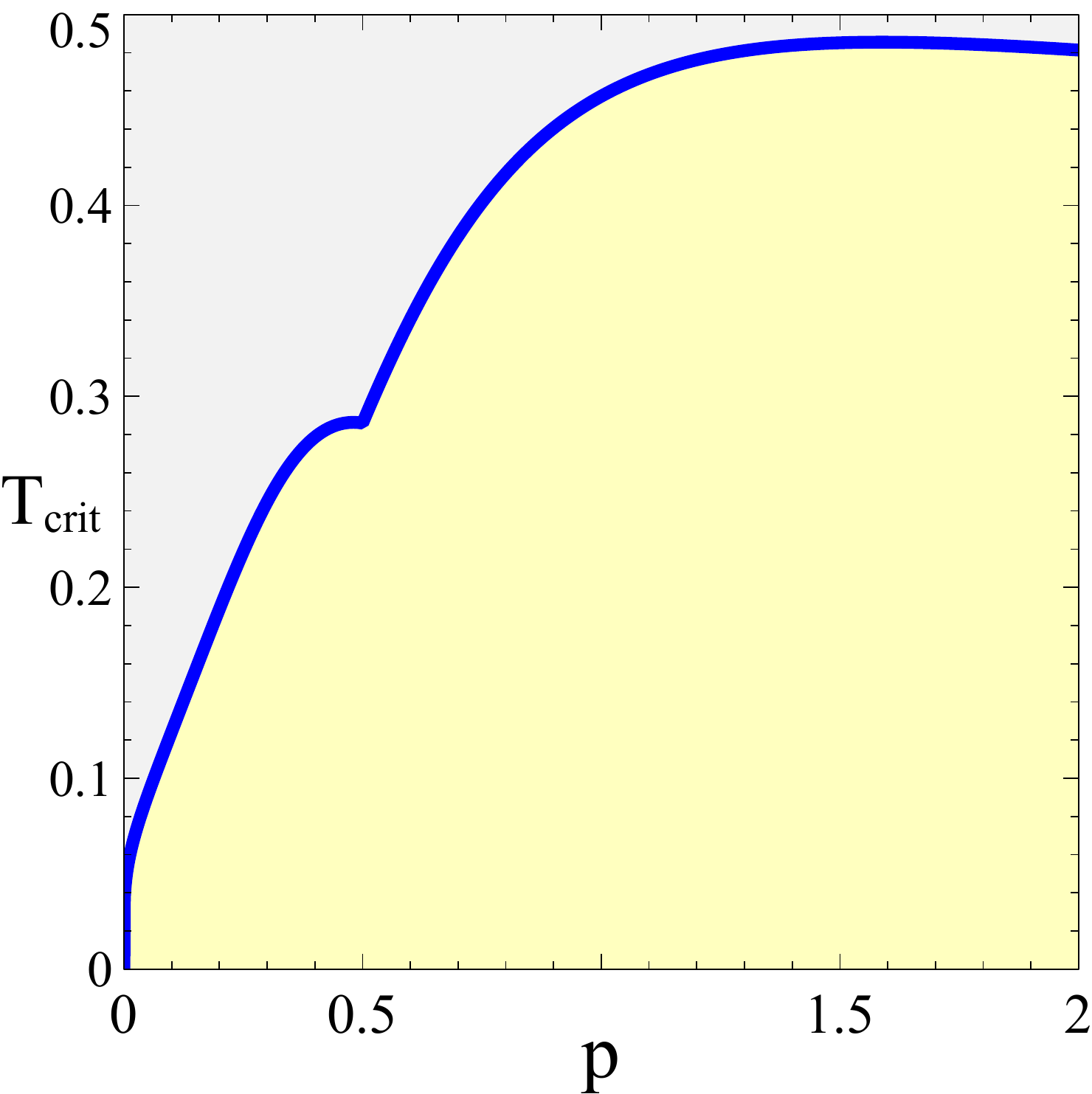} \quad
\includegraphics[width = 0.23\textwidth, keepaspectratio]{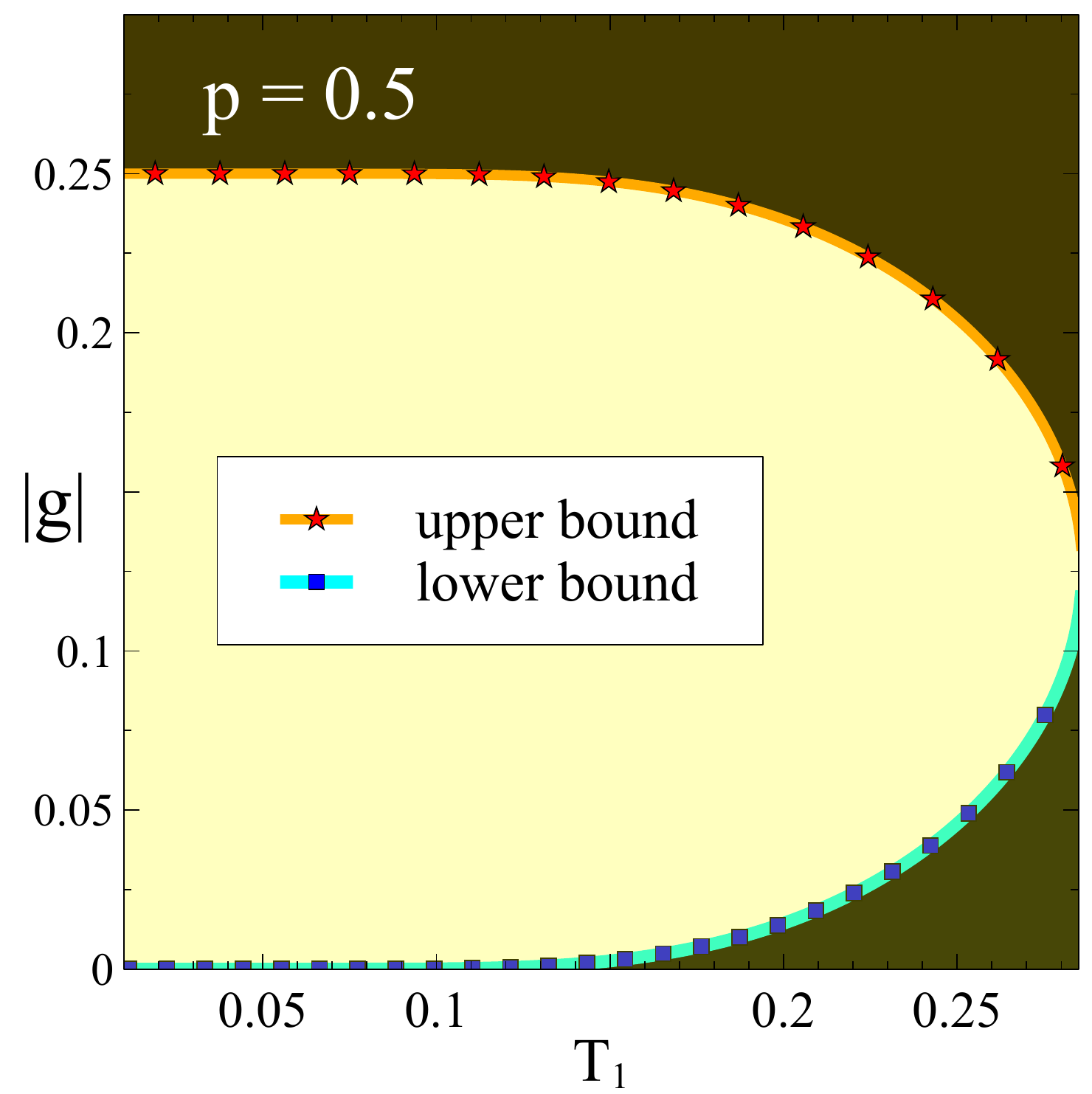} \quad
\caption{ Magic creation in the limit of low spin bath temperature $T_2$.   Left : dependence of critical temperature $T_{\text{crit}}$ on the reset probability $p_1 = p_2 = p$. Creation of magic is possible in the pale yellow region and impossible in the light gray region.   Right : allowed interval for interaction strength  $g$  with respect to heat bath temperature $T_1$ for creation of magic. Creation of magic is possible in the pale yellow region and impossible in the dark brown region.}
\label{low_temp}
\label{magic}
\end{figure}

Fig. \ref{magic} illustrates that the critical temperature increases with the reset probability $p$. However, even if the temperature of the heat bath is less than $T_{\text{crit}}$, the interaction strength $g$ must satisfy \eqref{condition_magic} or similar conditions for $r_x - r_y + r_z$ or $r_y - r_x + r_z$ for creation of magic. This effect is demonstrated in Fig. \ref{low_temp}, which shows that the allowed range of interaction strength $g$ steady decreases until it vanishes at the critical temperature $T_{\text{crit} }$.

 \begin{figure}
\includegraphics[width = 0.23\textwidth, keepaspectratio]{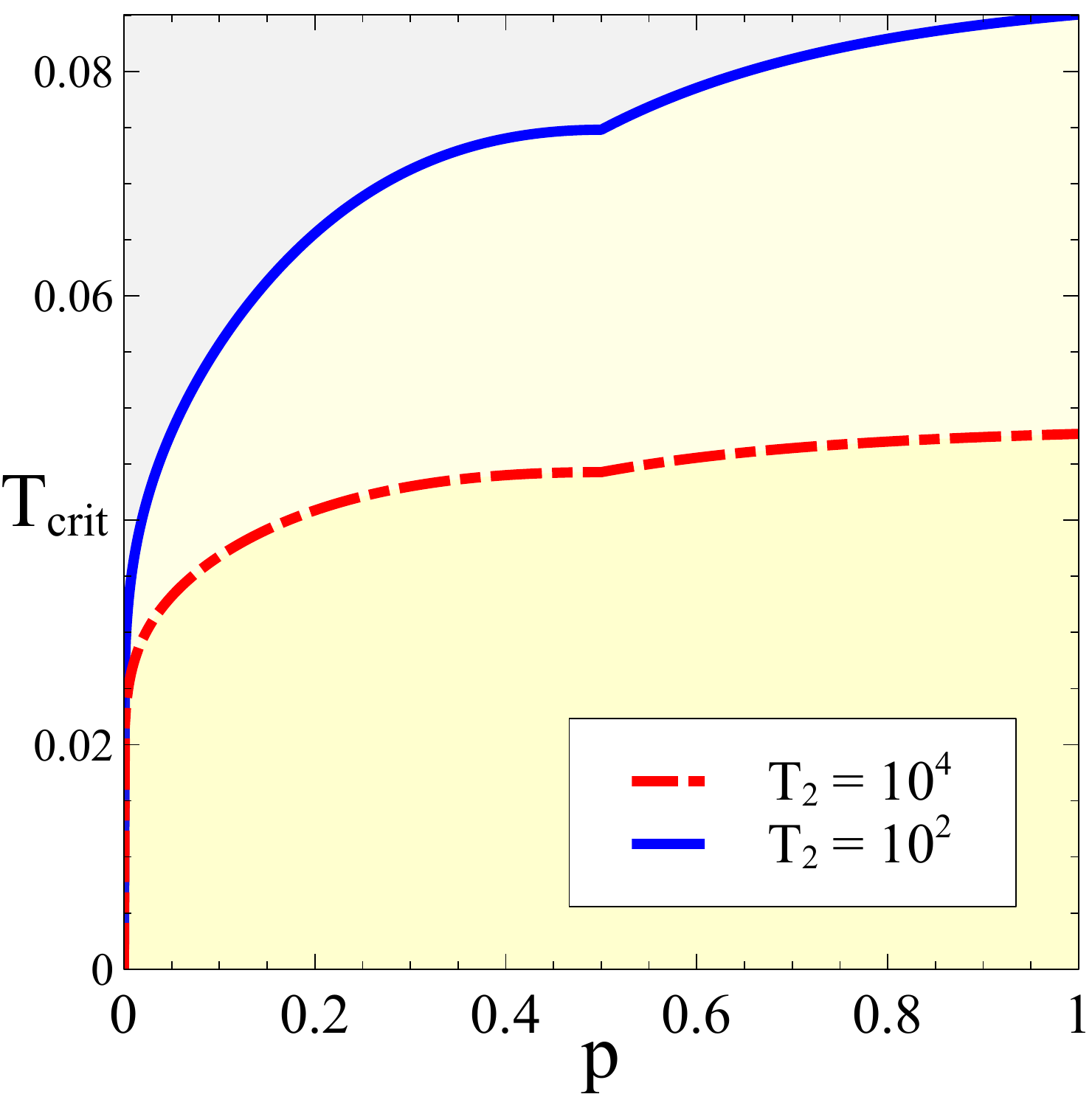} \quad
\includegraphics[width = 0.23\textwidth, keepaspectratio]{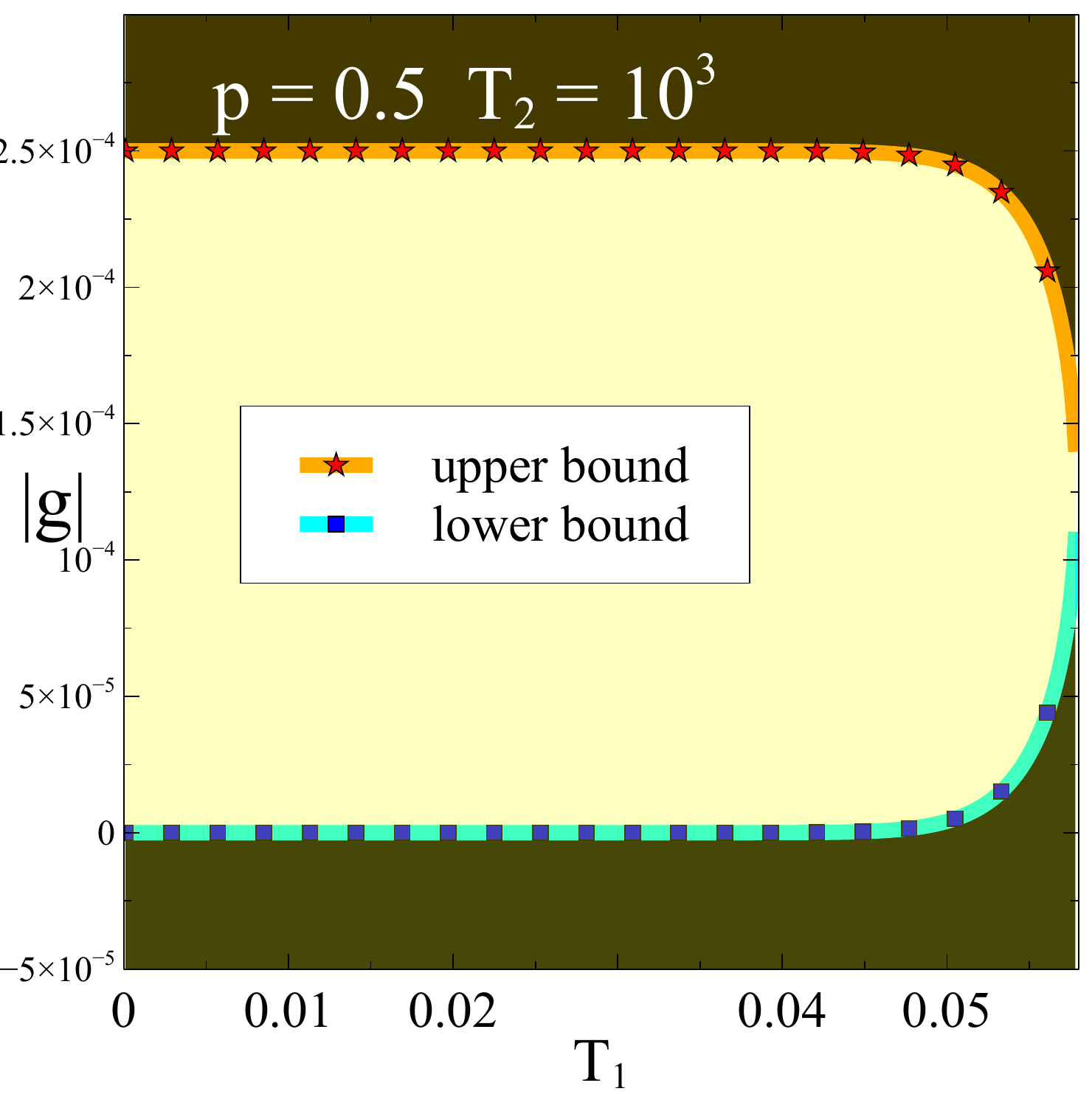} \quad
\caption{ Condition for creation of magic for high spin bath temperature $T_2$.   Left : dependence of critical temperature $T_{\text{crit}}$ on the reset probability $p_1 = p_2 = p$. Creation of magic is possible in the pale yellow region and impossible in the light gray region.  Right : Allowed interval for interaction strength  $g$  with respect to heat bath temperature $T_1$ for creation of magic. Creation of magic is possible in the pale yellow region and impossible in the dark brown region. }
\label{magic_high}
\end{figure}

Let us now explore the opposite limit, that is, the spin bath temperature $T_2$ being very high, and again make the simplifying assumption that $p_1 = p_2 = p$. We recall that, for $x \rightarrow \infty$, $\tanh (1/x) \approx 1/x $, and $\cosh(1/x) \approx 1$. Making these approximations yield the following result
\begin{align}
& r_x + r_y + r_z \approx \tanh \left(\frac{1}{2 T_1} \right) \left[1 + 2 g \frac{p(4p^2 + 4p -1)}{ T_{2}(1 + 4 p^2)^2 } -  \frac{g^2}{p^2} \right]   
\end{align}

From the above formula, following the approach earlier, the condition that $r_x + r_y + r_z > 1$ can be shown to be equivalent to  \beq \left( g - \frac{F_1}{F_2} \right)^2 < \  \frac{F_1^2}{F_2^2} - \frac{\lambda}{F_2}, \label{condition_magic_2} \eeq  where $F_1 = \frac{p(4 p^2 + 4 p -1)}{T_2 (1+4p^2)^2}$, and $F_2 = 1/p^2$. Similar to before, the critical threshold temperature $T_{\text{crit}}$ of the heat bath is the maximum of the critical threshold temperatures corresponding to the conditions for $r_x \pm r_y + r_z$, or $r_y - r_x + r_z$ surpassing unity respectively. That is, \beq T_{\text{crit}} = \max \left[\frac{1}{\ln \left(1 + \frac{F_2}{F_1^2} \right)}, \frac{1}{\ln \left(1 + \frac{F_2}{G_1^2} \right)}, \frac{1}{\ln \left(1 + \frac{F_2}{H_1^2} \right)} \right], \eeq where $G_1 =  \frac{p(1 + 4p -4 p^2)}{T_2 (1+4p^2)^2} $, and  $H_1 =  \frac{p(4 p^2 - 4p -1)}{T_2 (1+4p^2)^2}.$ In case the temperature of the heat bath is less than the critical temperature, the interaction strength $g$ must again satisfy either \eqref{condition_magic_2} or its analogues.  The above situations are pictorially depicted in Fig \ref{magic_high} from which we observe that the critical heat bath temperature for creation of magic is enhanced if the spin bath temperature is lowered. From Fig \ref{magic_high}, we also affirm that similar to the low temperature case, the window of interaction strength $g$ for which magic creation is possible becomes narrower and narrower with increasing heat bath temperature $T_1$ until vanishing when the heat bath temperature exceeds the critical temperature $T_{\text{crit}}$. In line with our naive expectation that it becomes harder and harder to extract quantumness from a system in presence of large classical noise, Fig \ref{magic_high} illustrates that for increased spin bath temperature $T_2$, the critical temperature of the thermal bath for creation of magic is significantly depressed.

\begin{figure}
\includegraphics[width = 0.22\textwidth, keepaspectratio]{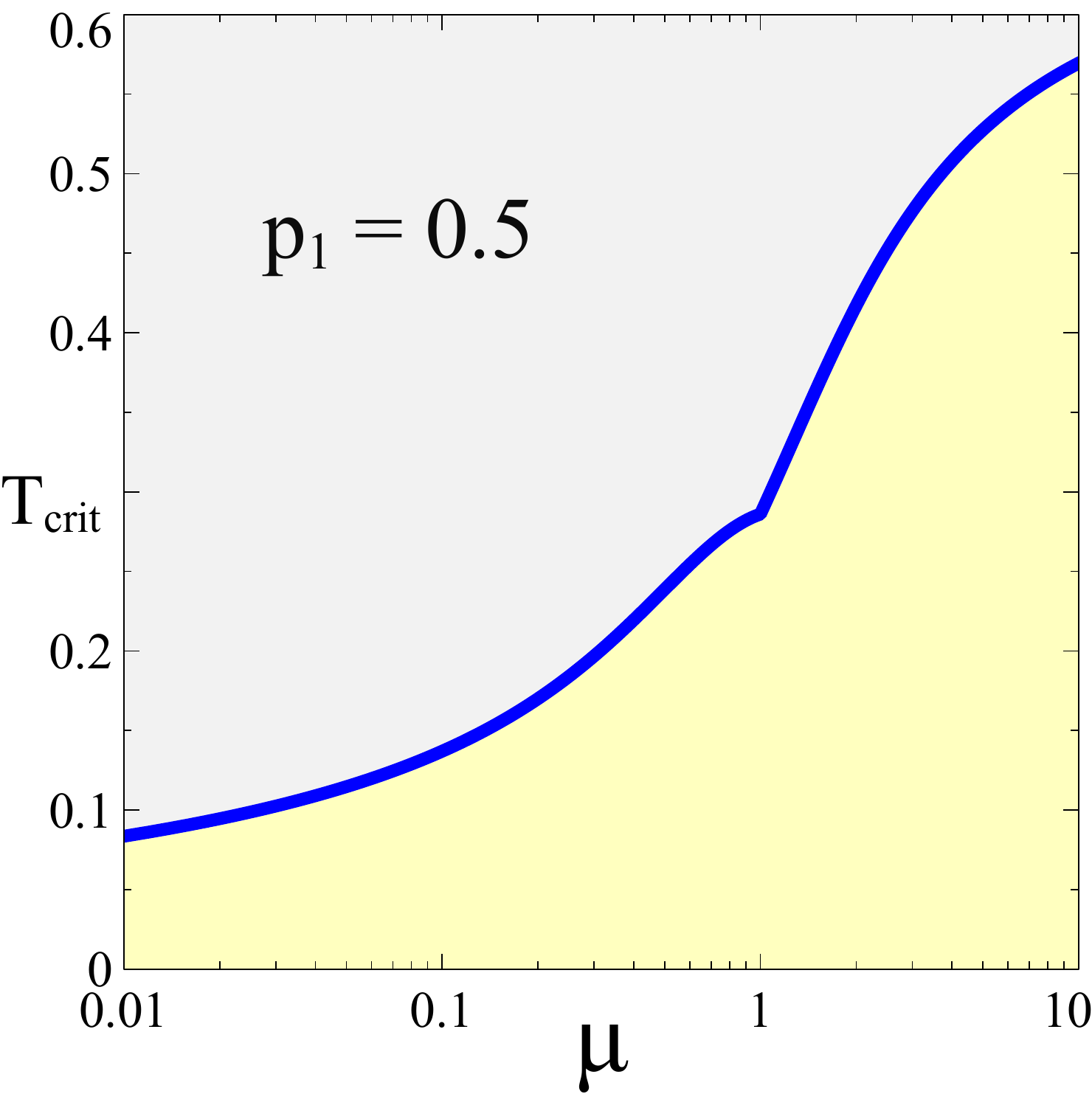} \quad
\includegraphics[width = 0.22\textwidth, keepaspectratio]{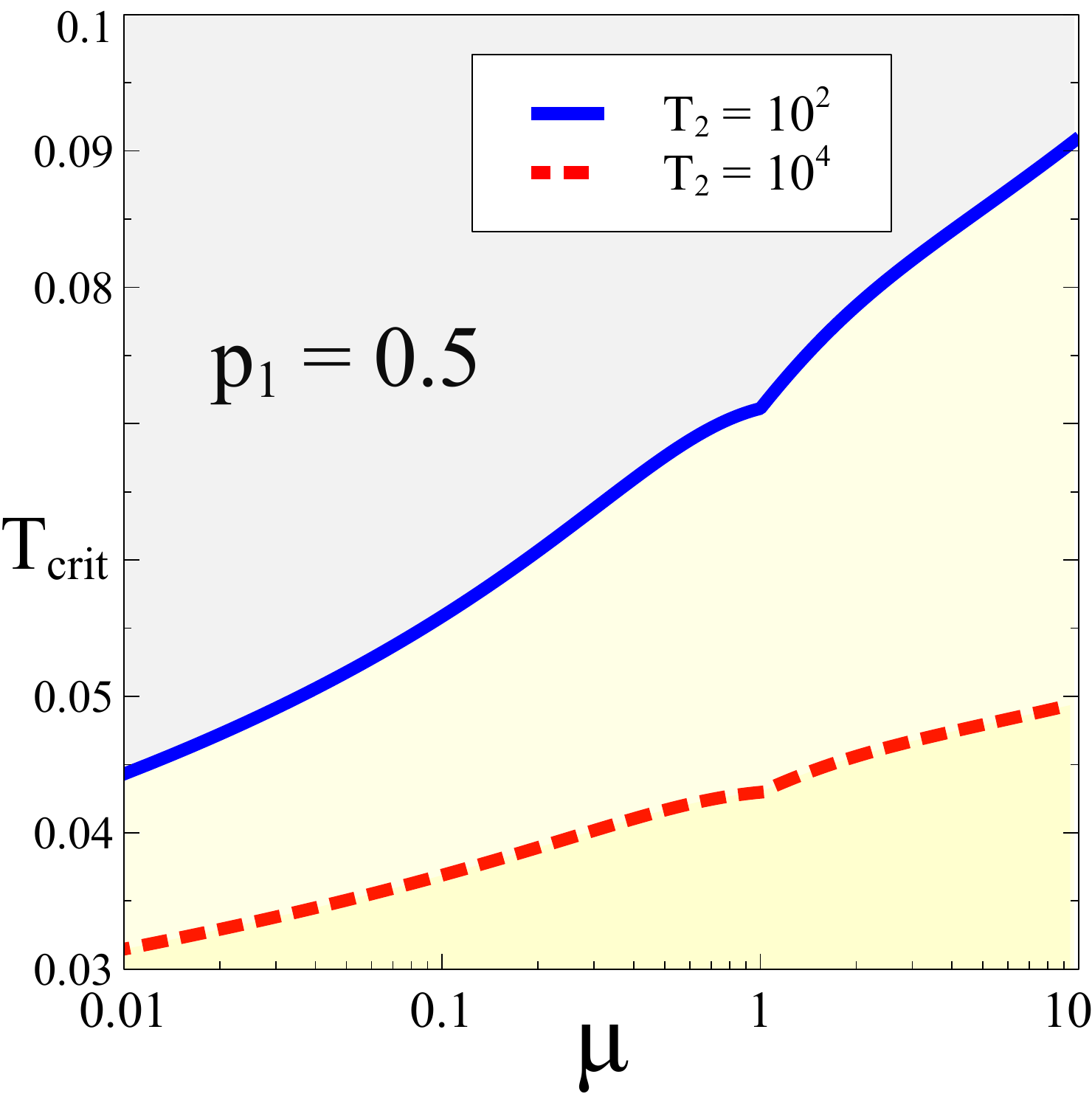} \quad
 \caption{Effect of asymmetry $\mu$ between the reset probabilities $p_1$ and $p_2 = \mu p_1$ on the critical temperature $T_{\text{crit}}$ for creation of magic in the low spin bath temperature $T_2$ limit (left) and high spin bath temperature $T_2$ limit (right). Creation of magic is possible in the pale yellow region and impossible in the light gray region.   }
\label{magic_asymm}
\end{figure}

Until now, we have made the simplifying assumption that the reset probabilities are equal. Let us now study, in the low $T_{2}$ limit, the effect of assymetry between the reset probabilities. Suppose $p_1 = p$ and $p_2 = \mu p$. Thus, in the low $T_2$ limit, the corresponding expressions are  
\begin{widetext}
\begin{align}
r_x + r_y + r_z  \approx \tanh \left(\frac{1}{2 T_1} \right) \left[1 + 4 g \mu p \frac{4\mu p^2 + 2 \mu p + 2 p -1}{ (1 + 4 p^2)(1 + 4 \mu^2 p^2) } -  2 g^2 \frac{1 + 2\mu^2 p^2 + 4 p^2 + 16 \mu^2 p^4 + 8 \mu^3 p^4}{p^2 (1+\mu)(1 + 4 p^2)(1 + 4 \mu^2 p^2)}  \right]   \\
r_x - r_y + r_z   \approx \tanh \left(\frac{1}{2 T_1} \right) \left[1 + 4 g \mu p \frac{1 + 2 \mu p + 2 p -4\mu p^2 }{ (1 + 4 p^2)(1 + 4 \mu^2 p^2) } -  2 g^2 \frac{1 + 2\mu^2 p^2 + 4 p^2 + 16 \mu^2 p^4 + 8 \mu^3 p^4}{p^2 (1+\mu)(1 + 4 p^2)(1 + 4 \mu^2 p^2)}  \right]   \\
-r_x + r_y + r_z  \approx \tanh \left(\frac{1}{2 T_1} \right) \left[1 + 4 g \mu p \frac{4\mu p^2 - 2 \mu p - 2 p -1}{ (1 + 4 p^2)(1 + 4 \mu^2 p^2) } -  2 g^2 \frac{1 + 2\mu^2 p^2 + 4 p^2 + 16 \mu^2 p^4 + 8 \mu^3 p^4}{p^2 (1+\mu)(1 + 4 p^2)(1 + 4 \mu^2 p^2)}  \right]   
\end{align}
\end{widetext}
In the opposite, i.e., high $T_2$ limit, using previously stated approximations, viz., $\tanh(1/x) \approx 1/x$ and $\cosh(1/x) \approx 1$ for $x \rightarrow \infty$, the expressions for linear functions of Bloch vectors are given by 
\begin{widetext}
\begin{align}
r_x + r_y + r_z  \approx \tanh \left(\frac{1}{2 T_1} \right) \left[1 + 2 g \mu p \frac{4\mu p^2 + 2 \mu p + 2 p -1}{T_2 (1 + 4 p^2)(1 + 4 \mu^2 p^2) } -  2 g^2 \frac{1}{p^2 (1+\mu)}  \right]   \\
r_x - r_y + r_z   \approx \tanh \left(\frac{1}{2 T_1} \right) \left[1 + 2 g \mu p \frac{1 + 2 \mu p + 2 p -4\mu p^2 }{ T_2 (1 + 4 p^2)(1 + 4 \mu^2 p^2) } -  2 g^2 \frac{1}{p^2 (1+\mu)}   \right]   \\
-r_x + r_y + r_z  \approx \tanh \left(\frac{1}{2 T_1} \right) \left[1 + 2 g \mu p \frac{4\mu p^2 - 2 \mu p - 2 p -1}{T_2 (1 + 4 p^2)(1 + 4 \mu^2 p^2) } -   2 g^2 \frac{1}{p^2 (1+\mu)}    \right]   
\end{align}
\end{widetext}
Fig \ref{magic_asymm} illustrates that, in both the low $T_2$ and high $T_2$ limit, the larger the reset probability of the spin bath is compared with te reset probability of the heat bath, the more the magnitude of critical temperature for creation of magic.  

However, when considering these results, one must also keep in mind that they have been obtained through a perturbation expansion in $g$. Thus, the cases where magic creation seems possible from the relations like \eqref{condition_magic} or \eqref{condition_magic_2}, yet the interaction strength is quite high, have to be more carefully treated. Moreover, if the interaction strength is quite high, the reset model master equation itself may not work.

\section*{CONCLUDING REMARKS}
In this work, we have proposed an autonomous system of two qubits attached respectively to a heat bath and an angular momentum bath and interacting through an energy-exchange Hamiltonian and shown that the reduced qubit attached to the heat bath may have certain quantum features like coherence and magic even at the steady state. We have demonstrated the existence of a critical threshold temperature of the heat bath, above which, creation of magic is not possible. Even below the critical temperature, we have observed that there is an allowed range for the strength of interaction, if magic has to be created in the reduced qubit. 

Presently, we have only looked at the condition of existence of magic. In future  work, it may be useful to investigate the $quantity$ of magic created. Since the smallest quantum system for which we currently have an exact and analytically computable measure of magic is a qutrit (via the negativity of discrete Wigner functions),  replacing our two qubit model with a qutrit-qubit or a two qutrit model may be a possible approach. It may also be useful to consider whether such angular momentum baths offer any advantage over a heat bath in usual tricycle type absorption refrigerator models. In the context of the present work, the reset based master equation used here is quite simplistic and it would be nice to extend the present study to analyze the dynamics explicitly  for specific concrete bath models through  Lindblad-type master equations. As has been observed in  the performance of quantum absorption refrigerators, it would be interesting to investigate whether the transient performance, i.e., coherence and magic generated in the transient state, may exceed the steady coherence and magic. Self-contained quantum absorption refrigerators are practically relevant for building  quantum computers, since cooling gets rid of thermal noise, which  allows one to freely concentrate on correcting quantum fluctuation induced errors. In this work, we have concentrated on generation of magic, an ingredient of non-classical gate implementation. Whether there is a trade-off between generation of magic and cooling rate for suitably designed quantum absorption refrigerators may be an interesting avenue to explore. 
\\

I thank Sk Sazim and Shiladitya Mal for discussions and acknowledge a graduate fellowship from Department of Atomic Energy, Govt. of India.
\bibliography{solo_ref} 
\bibliographystyle{apsrev4-1}
\end{document}